\documentclass[aps,prb,preprint]{revtex4}
\usepackage[version=3]{mhchem} 
\usepackage{rotating}
\usepackage{subfig}

\begin{document}
\title{Molecular explanation for why talc surfaces can be both hydrophilic and hydrophobic}

\author{Benjamin Rotenberg}
\affiliation
{CNRS et UPMC-Paris6, Laboratoire PECSA, UMR 7195, 4 pl. Jussieu, F-75005 Paris, France}
\email{benjamin.rotenberg@upmc.fr}
\author{Amish J. Patel}
\affiliation
{Howard P. Isermann Department of Chemical \& Biological Engineering, and Center for Biotechnology and Interdisciplinary Studies, Rensselaer Polytechnic Institute, Troy, New York 12180}
\author{David Chandler}
\affiliation
{Department of Chemistry,
University of California, Berkeley, California 94720}

\date{\today}



\newpage
\begin{abstract}
While individual water molecules adsorb strongly on a talc surface (hydrophilic behavior), a droplet of water beads up on the same surface (hydrophobic behavior).
 To rationalize this dichotomy, we investigate the influence of the microscopic structure of the surface and the strength of adhesive (surface-water) interactions on surface hydrophobicity. 
We show that at low relative humidity, the competition between adhesion and the favorable entropy of being in the vapor phase determines the surface coverage. 
However, at saturation, it is the competition between adhesion and cohesion (water-water interactions) that determines surface hydrophobicity.
The adhesive interactions in talc are strong enough to overcome the unfavorable entropy, and water adsorbs strongly on talc surfaces. However, they are too weak to overcome the cohesive interactions, and water thus beads up on talc surfaces.
Surprisingly, even (talc-like) surfaces that are highly adhesive, do not fully wet at saturation. Instead, a water droplet forms on top of a strongly adsorbed monolayer of water.
Our results imply that the interior of hydrophobic zeolites suspended in water may contain adsorbed water molecules at pressures much smaller than the intrusion pressure.
\end{abstract}

\maketitle

\newpage
\subsection{Introduction}

Wetting properties of minerals in soils and rocks play a
crucial role in the transport, and thus availability, of water and oil.
Clay minerals are particularly interesting, not only due
to their abundance in nature and in synthetic
materials, but also because the existence of clays with different structures allows us to investigate the effect of surface microstructure on macroscopic properties.
Clay surfaces can be either charge-neutral or have a net charge, which is balanced by counter-ions in solution. Molecular simulation has furthered our understanding of both these types of clays: uncharged clays have been studied using both ab-initio~\cite{Bridgeman_MolPhys_1996,Churakov07} and classical simulations~\cite{Cygan_JPCB_2004,Cygan_JMaterChem_2009}, whereas simulations of charged clays have provided insights into interlayer properties~\cite{Delville91,Marry03,BoekLiNaK95,Sposito99}, 
swelling~\cite{DelvilleNa92,Young00,Hensen02,Tambach04a},
and cation exchange~\cite{Teppen,EdgesGCA07,RotenbergGCA09}.
These studies have shown that the surface microstructure is expected to be more important in determining surface-water interactions in uncharged clays~\cite{Wang05,Marry_PCCP_2008}, and it is these surfaces that are the focus of the current work.
Among uncharged clays, talc surfaces have attracted a lot of 
attention~\cite{Wang_EPSL_2004,Wang_GCA_2006,Wang_JPhysChemC_2009},
because of their peculiar behavior with respect to water.
Water adsorption at low relative humidity (RH) reveals the
presence of strong binding sites on talc~\cite{Michot_Langmuir_1994}. 
Such strong binding sites are absent in other uncharged clays such as 
pyrophyllite and fluorotalc. Yet, experimental contact angles indicate that the surface of  talc monocrystals is hydrophobic, similar to that of 
pyrophyllite~\cite{Giese_PhysChemMinerals_1991,VanOss_ClaysClayMiner_1995}.

To investigate this dichotomy, here we employ molecular dynamics simulations combined with recently developed algorithms \cite{Patel_JPhysChemB_2010,AJP_JStatPhys}.
In agreement with experiments, we find that at low RH, 
talc surfaces display hydrophilic behavior as water adsorbs strongly 
to the binding sites on the surface. 
However, at saturation, cohesive interactions dominate and the interaction between the surface binding sites and water is minimal, resulting in a hydrophobic surface.

To further explore the role of surface microstructure and the strength of the adhesive interactions on surface hydrophobicity, we also study similar clay minerals, pyrophyllite and fluorotalc, as well as modified talc surfaces with a range of binding site polarities, both at low relative humidity and at saturation.
We find that the dual hydrophilic-hydrophobic behavior observed in talc, is generically expected to manifest for surfaces whose adhesive interaction energy lies in a special range. 
If the adhesion to water is strong enough to overcome the entropy of being in the vapor phase at low RH, water adsorbs strongly to the surface (hydrophilic behavior). At the same time, if adhesion is too weak to overcome the cohesive interactions in water, the surface is hydrophobic at saturation.
 For modified talc surfaces with strong enough adhesion to overcome the cohesive interactions, all surface binding sites are occupied by water molecules at saturation, as expected. Surprisingly, instead of observing complete wetting, we find that a water droplet sits atop the adsorbed water monolayer.

\subsection{Microscopic Models}
\label{sec:Clays}

\begin{figure}[t!]
\begin{center}
\hspace{-0.17in}\includegraphics[width=3.50in]{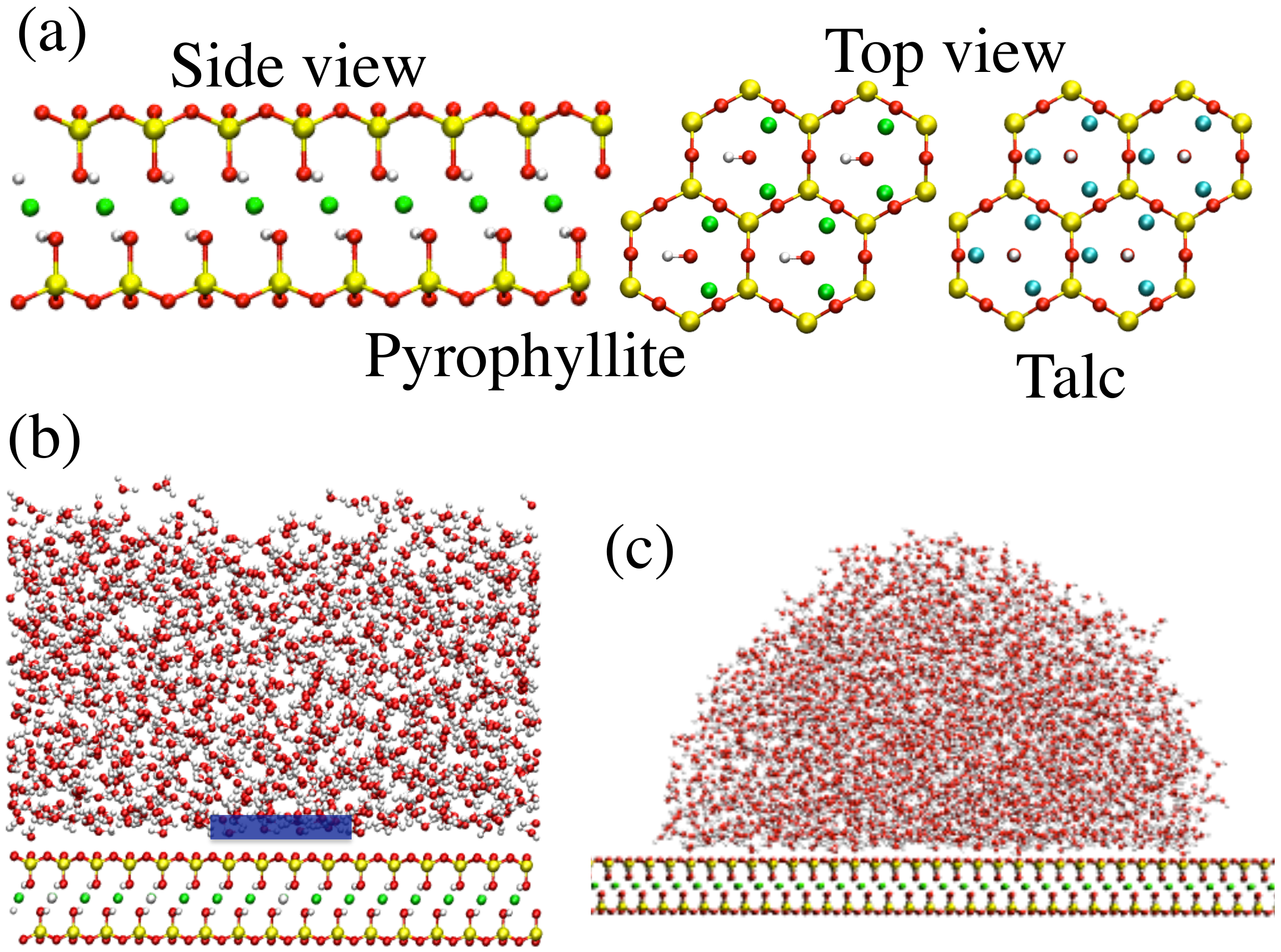}
\end{center}
\caption{\label{fig:structure}
(a) Microscopic clay structure (Red: O, White: H, Yellow: Si, Green: Al, Cyan: Mg atoms). 
The side and top views of the pyrophyllite clay sheet show the hydroxyl 
(-OH) groups that are parallel to the sheet. 
In talc (top view shown), the -OH groups are perpendicular to the sheet 
and can participate in hydrogen bonds with water.
In fluorotalc (not shown), the talc -OH groups are replaced by F atoms. 
(b) Part of the simulation setup for studying the clay - water interface.
The blue box is the observation volume, ${\rm v}$, used to probe density fluctuations.
(c) Simulation setup for determining contact angles.
}
\end{figure}

Talc, fluorotalc and pyrophyllite are uncharged clay minerals, {\it i.e.}, layered silicates of magnesium (Mg) or aluminum (Al). They belong to the family of TOT clays: each clay sheet consists of a layer of octahedrally
coordinated Mg or Al oxide between two layers of tetrahedral
silicon oxide (see~\ref{fig:structure}(a) - side view). 
The surface of these sheets displays hexagonal rings of SiO$_2$ tetrahedra.
In talc and fluorotalc, all octahedral sites are occupied by Mg atoms,
while in pyrophyllite two third of these sites are occupied by Al atoms
(see~\ref{fig:structure}(a) - top view).
The charge on Mg and Al is balanced by hydroxyl groups in the center of the 
hexagonal cavities. In talc, these hydroxyl groups are oriented perpendicular 
to the surface, and can participate in hydrogen bonds with water.
In pyrophyllite, the hydroxyl groups are oriented parallel to the surface, and in fluorotalc, they are replaced by fluorine atoms. The atomic coordinates for the unit cells of these clays have been included as Supplementary Information.

We use the CLAYFF force field~\cite{Cygan_JPCB_2004} to model the interactions 
of the clay atoms and the SPC/E model to describe water~\cite{Berendsen_JPC_1987}. Lorentz-Berthelot combination rules are used to determine the pair Lennard-Jones parameters and a rigid clay structure is assumed. 
As there are no parameters for fluorine in CLAYFF, we assigned it a charge 
equal to that of the -OH group in talc (-0.525) and
Lennard-Jones parameters of the fluoride ion reported in Ref.~\cite{Dang_CPL_1992}.
All simulations were performed in the NVT ensemble using the LAMMPS simulation 
package~\cite{LAMMPS} at a temperature, $T=300$~K, maintained using a Nose-Hoover
thermostat~\cite{Martyna92}. SHAKE was used to integrate the motion of the 
rigid water molecules~\cite{Ryckaert77} and long-range electrostatic interactions were computed using Ewald summation.

\subsection{Methods}
\label{sec:Methods}
\subsubsection{Clay - water interface}
A clay-water interface is representative of the situation at saturation.
The setup shown in~\ref{fig:structure}(b) is used to calculate the local water density, 
$\rho(z)$, as well as the water density fluctuations near the interface.
The potential of mean force, $\mathcal{F}$, for bringing a water molecule from bulk to a distance $z$ from the plane of the Mg atoms for talc and fluorotalc 
(and Al for pyrophyllite) is related to $\rho(z)$ by $\mathcal{F}(z)=-k_{\rm B}T\ln[\rho(z)/\rho_{\rm b}]$, where $k_{\rm B}$ is the Boltzmann constant and $\rho_{\rm b}$ is the bulk water density.
To quantify density fluctuations, we measure the probability distribution, $P_{\rm v}(N)$, of finding $N$ water molecules in an observation volume ${\rm v}$, adjacent to the clay surface, using the indirect umbrella sampling (INDUS) method~\cite{Patel_JPhysChemB_2010,AJP_JStatPhys}. 
We chose a rectangular parallelopiped of dimensions 
$15\times 15\times 3$~\AA$^3$ placed near the surface [see~\ref{fig:structure}(b)], as the observation volume. The exact $z$-position of ${\rm v}$ was chosen so that
the mean water density in ${\rm v}$ is equal to $\rho_{\rm b}$.
The simulation box also contained a fixed wall of repulsive WCA particles (not shown), placed at the top of the box (far from ${\rm v}$) to nucleate a vapor-liquid buffering interface.
 \\
\subsubsection{Contact angle}
The simulation setup for contact angle measurements is shown in~\ref{fig:structure}(c). The contact angle is determined by computing water density maps
in the plane of the center-of-mass of the drop.
The curve with density equal to half of the bulk density is then fit to a circle and the 
angle between the tangent to this circle at $z_S=7$~\AA\ and the horizontal axis is 
taken to be the contact angle. While the exact quantitative value of the contact 
angle depends on the choice of $z_S$, our qualitative findings do not. \\
\subsubsection{Water vapor adsorption}
The adsorption of water vapor at low RH corresponds the interaction of an isolated water molecule with the surface.
To determine the corresponding adsorption free energy, $\Delta\mu_{\rm ads}$, we compute $\mathcal{F}(z)$ using umbrella sampling, with the weighted histogram analysis method (WHAM)
~\cite{Kumar_JCompChem_1995,Roux_CompPhysComm_1995} being used to reconstruct $\mathcal{F}(z)$ from the biased trajectories. \\

\subsection{Hydrophobicity at low and high RH}
\label{sec:PMF}
Using the various molecular measures of hydrophobicity described above, we study talc, as well as fluorotalc and pyrophyllite surfaces, both at saturation and at low RH. \\
\subsubsection{High RH}
\begin{figure*}[ht!]
\centering
\hspace{-0.in}\includegraphics[width=7.in]{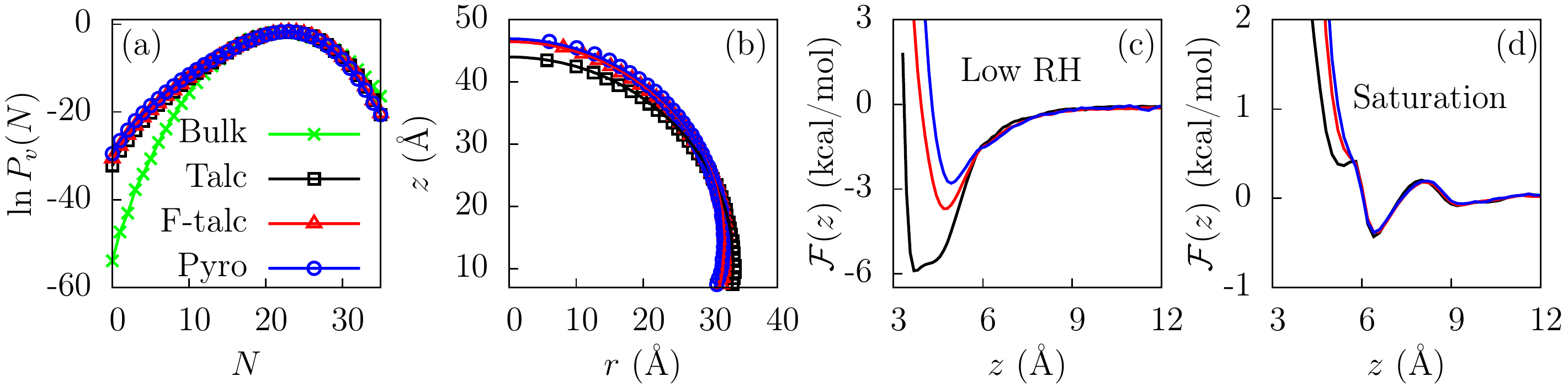}
\caption{
(a) The probability, $P_{\rm v}(N)$, of observing $N$ water molecules in a probe 
volume (${\rm v}=15\times 15\times 3$~\AA$^3$) displays a low $N$ fat tail when ${\rm v}$ is near the surface of talc (black), fluorotalc (red), and pyrophyllite (blue), as compared to that when ${\rm v}$ is in bulk water (green).
(b) Water droplet profiles corresponding to $\rho(r,z)=0.5\rho_{\rm b}$ are 
shown for the clay surfaces.
The contact angles for the surfaces are similar: $96^{\circ}$ for talc, $103^
{\circ}$ for fluorotalc, and $105^{\circ}$ for pyrophyllite (based on tangents 
drawn at $z_S=7$\AA).  
(c) Potential of mean force, $\mathcal{F}(z)$, for the adsorption of an 
isolated water molecule (low RH) to the clay surfaces.
The hydrogen atoms of the talc -OH groups are located at $z=2$~\AA\ and 
can participate in hydrogen bonds with water molecules. 
(d) $\mathcal{F}(z)$ at the clay - liquid water interface (saturation). 
To maximize H-bonding with other waters, the binding site is no longer occupied.
}
\label{fig:pmf}
\end{figure*}

Theory~\cite{Lum_JPhysChemB_1999,Chandler_Nature_2005,Berne_Weeks_rev,Varilly_JChemPhys_2011} and simulations~\cite{Mittal_PNAS_2008,Sarupria_PhysRevLett_2009,Godawat_PNAS_2009,Patel_JPhysChemB_2010,Acharya_Faraday} have shown that the mean water density near a surface is not a good measure of its hydrophobicity.
Instead, fluctuations away from the mean, and in particular, the rare 
fluctuations~\cite{Patel_JPhysChemB_2010} indicating the cost of creating a 
cavity at the interface correlate quantitatively with the contact 
angle~\cite{AJP_Lscale}.
Patel {\it et al.} have shown that hydrophobic surfaces display an enhanced 
probability of density depletion or a low $N$ fat tail in the $P_{\rm v}(N)$ 
distribution, while $P_{\rm v}(N)$ near hydrophilic surfaces is similar to that in bulk water~\cite{Patel_JPhysChemB_2010}. 
As shown in~\ref{fig:pmf}(a), $P_{\rm v}(N)$ near all three clay surfaces displays 
a low $N$ fat tail, indicating that these surfaces are hydrophobic. A slight lifting of 
the fat tail from talc to fluorotalc and pyrophyllite suggests a corresponding 
marginal increase in hydrophobicity.

Another way to probe surface hydrophobicity is by simulating a sufficiently large water droplet on the surface and estimating the corresponding contact angle.
\ref{fig:pmf}(b) shows the average shape of droplets on the clay surfaces. 
The curve corresponding to $\rho(r,z)=0.5\rho_{\rm b}$ 
is a circle in the ($r,z$) plane, where $r$ is the distance from the axis that 
passes through the center of mass of the droplet.
The contact angles obtained by tangents drawn at $z_S=7$\AA~on the three surfaces 
are similar (talc: $96^{\circ}$, fluorotalc: $103^{\circ}$ pyrophyllite: $105^{\circ}$), and clearly indicate hydrophobic behavior.

Reliable experimental estimates of the contact angle of water droplets on both talc 
and pyrophyllite monocrystals are between 80$^\circ$ and 85$^\circ$~\cite{Giese_PhysChemMinerals_1991,VanOss_ClaysClayMiner_1995}.
The reported values for measurements on powders are usually smaller due to the 
presence of hydrophilic sites on the edges of finite clay particles~\cite{Douillard_JCIS_2002}. 
To the best of our knowledge, no experimental contact angles have been reported for fluorotalc. 
For both talc and pyrophyllite, the contact angles obtained from our simulations 
($96^{\circ}$ and $105^{\circ}$ respectively) are somewhat larger than the 
experimental estimates, suggesting that surfaces modeled with the CLAYFF model 
are too hydrophobic. Nevertheless, amongst various commonly used clay 
force fields~\cite{Skipper89,Smith98,Heinz_ChemMater_2005}, we find that the 
correspondence with experiments is closest for CLAYFF.  A comparison of these 
force fields with experiments is provided in the Supplementary Information.\\
\subsubsection{Low RH}
To investigate the wetting behavior of clay surfaces at low RH, we calculate the 
potential of mean force, $\mathcal{F}(z)$, for the adsorption of an isolated water 
molecule. $\mathcal{F}(z)$ displays a minimum near all the clay surfaces [see~\ref{fig:pmf}(c)], corresponding to an adsorption (or binding) free energy, $\Delta\mu_{\rm ads}$.
For talc, $\Delta\mu_{\rm ads}\approx-5.9$~kcal/mol, or $10~k_{\rm B}T$, 
consistent with the formation of a hydrogen bond between the water molecule and the hydroxyl group in talc. 
In fluorotalc, the hydroxyl group is replaced by fluorine, resulting in a reduction in 
$\Delta\mu_{\rm ads}$ to -3.5~kcal/mol. It also shifts the location of the minimum 
out by $\approx1$~\AA\ as the water is no longer strongly bound to the surface. 
Pyrophyllite, with the hydroxyl group parallel to the surface has an even
smaller $\Delta\mu_{\rm ads}\approx-2.8$~kcal/mol, and the minimum is
shifted out even more. 

To compare our estimate of $\Delta\mu_{\rm ads}$ from simulations 
to experimental data, we analyzed the data of Michot {\it et
al.}~\cite{Michot_Langmuir_1994} using a Langmuir model. 
This model assumes that there are no interactions between the adsorbed 
molecules and predicts a surface coverage, $\Theta = (P/P^*)/(1+P/P^*)$. 
$P^*$ is the pressure at which half of the surface sites are occupied 
and is related to $\Delta\mu_{\rm ads}$ through
\begin{equation}
\label{eq:pstar}
P^*=\frac{\sigma_{\rm max}k_{\rm B}T}{\delta}e^{\beta\Delta\mu_{\rm ads}},
\end{equation}
where $\sigma_{\rm max}\approx 4.2$~nm$^{-2}$ is the surface site density,
$\delta\approx 1-2$~\AA\ is the width of the surface layer, {\it i.e.} the width of the 
PMF well in~\ref{fig:pmf}(c), and $1/\beta=k_{\rm B}T$ is the thermal energy.

In the very low RH limit, corresponding to single water adsorption, 
we can safely assume that water molecules do not interact with each other. 
In this regime, $\Theta\approx P/P^*$ and the data in Figure 11 of Ref.~\cite{Michot_Langmuir_1994}, 
allow us to obtain an experimental estimate of $P^*\approx 0.056 P_{\rm sat}$ 
for the talc surface. Here, $P_{\rm sat}=30$~mbar is the saturation pressure of water. 
Using this value of $P^{*}$ in equation~\ref{eq:pstar}, 
we get an experimental estimate of  $\Delta\mu_{\rm ads}\approx-8$~kcal/mol~\cite{edges}. This somewhat stronger adsorption than that predicted from simulations 
using CLAYFF (-5.9~kcal/mol), is consistent with the overestimate of the CLAYFF talc contact angle.

If we further assume that the adsorbed water molecules do not interact 
with each other even at higher RH, the Langmuir model (with $P^*=0.056 P_{\rm sat}$) predicts that $\Theta\approx0.9$ at $50\%$ RH.
As water coverage on the talc surface can be large even 
at moderate RH, interactions between water molecules may be important, consistent with suggestions that clustering needs to be considered~\cite{Michot_Langmuir_1994,Carvalho_AppliedSurfaceScience_2007}. 
In contrast, for fluorotalc $\Theta$ at saturation estimated from $\Delta\mu_{\rm ads}$ is very small ($\approx1.5\%$), in agreement with the hydrophobic adsorption behavior reported in Figure 10 of Ref.~\cite{Michot_Langmuir_1994}. 

Thus, the clay surfaces simulated using the CLAYFF force field are more hydrophobic than the real clay surfaces used in experiments. However, the interesting dichotomy of talc surfaces is also observed in the simulations and our findings are qualitatively consistent with the experiments, both at low RH (strong adsorption for talc and not the other clays) and at high RH (large contact angles for all clays). 


\subsection{Cohesion vs Adhesion}
\label{sec:Discussion}

To investigate the disparate behavior of talc surfaces at low and high RH, 
we compare $\mathcal{F}(z)$ for moving a water molecule away from the surface under both conditions. 
At saturation, $\mathcal{F}(z)$ for the clay surfaces are similar [\ref{fig:pmf}(c)], consistent with similar droplet contact angle on the three surfaces [\ref{fig:pmf}(b)]. 
$\mathcal{F}(z)$ for fluorotalc is nearly identical to that for pyrophyllite, and that for talc features an additional local minimum around $z=5$~\AA\ corresponding to water molecules above the binding site.
However, the $\mathcal{F}(z)$ curves at saturation are qualitatively different 
from those at low RH [see \ref{fig:pmf}(c-d)]
For all three clays, and especially so for talc, the depth of the minimum at saturation is smaller than that at low RH, suggesting a weakening of adhesive interactions at saturation.

\begin{figure}[ht!]
\centering
\includegraphics[width=2.2in]{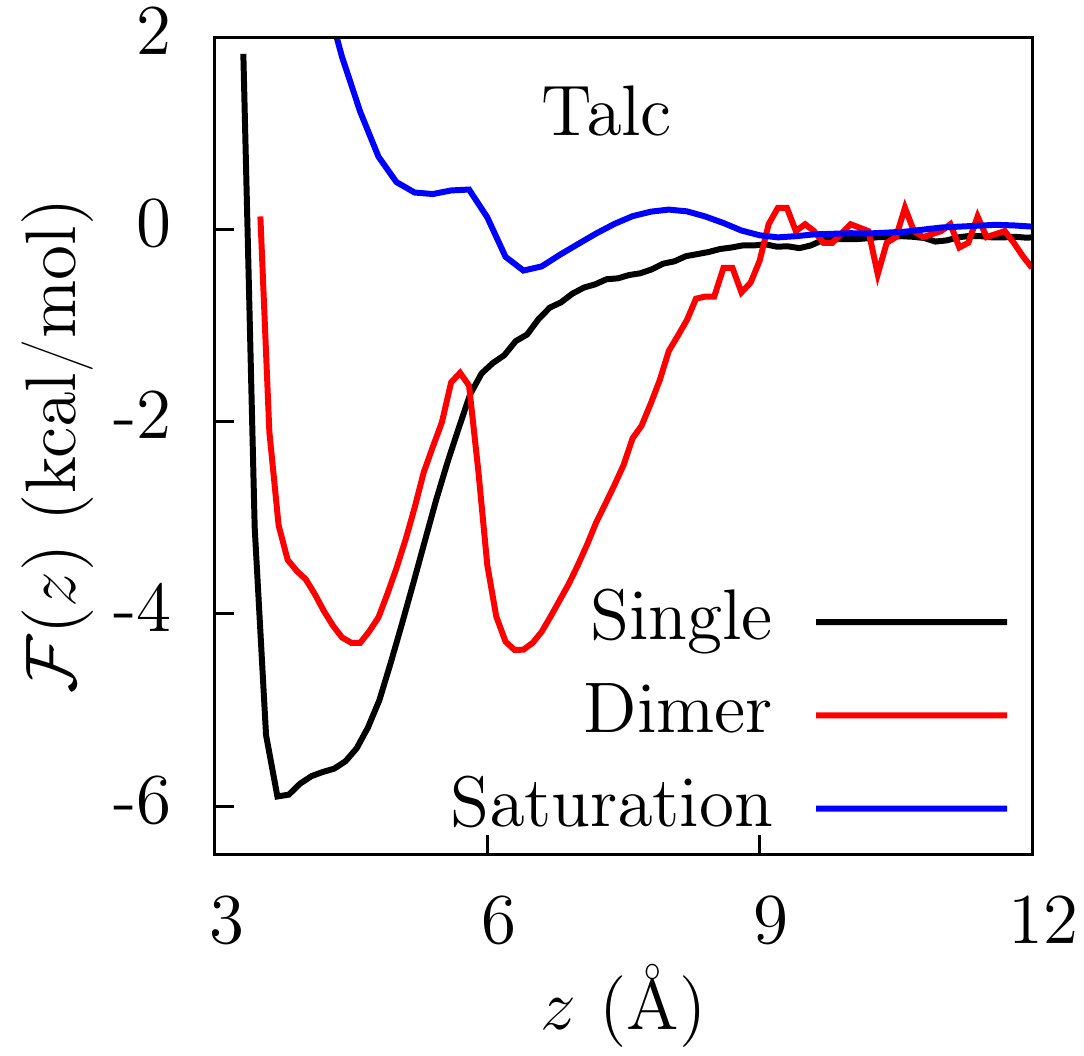}
\caption{
$\mathcal{F}(z)$ for adsorbing a single water molecule on the talc surface, compared to that for a molecule in the dimer and a molecule at saturation.
} 
\label{fig:dimer}
\end{figure}

To explore the competition between adhesive and cohesive interactions in talc, in \ref{fig:dimer}, we compare $\mathcal{F}(z)$ for an individually adsorbed water, with that for water in a dimer, and that for water at saturation.
As shown in \ref{fig:dimer}, the $\mathcal{F}(z)$ for the dimer displays two minima.
The minimum corresponding to the molecule inside the cavity is shifted to slightly larger values compared to the minimum in the $\mathcal{F}(z)$ for a single water. In addition, the depth of the minimum is smaller, and is comparable to that for a single water on the more hydrophobic fluorotalc surface [\ref{fig:pmf}(c) and \ref{fig:dimer}]. In other words, the presence of the second water weakens the adhesive surface-water interactions, which have to compete with the cohesive interactions between the waters.
As the dimer is less tightly bound to the surface than a single water, it is easier for the water to escape the cavity in the presence of a second molecule. 
The dimer is in fact more mobile on the talc surface than isolated water molecules (not shown), confirming that the interaction of the surface with the dimer is weaker than with individual molecules.
Finally, at saturation, cohesive interactions prevail, and water no longer occupies the binding site cavity as evidenced by the lack of a minimum in $\mathcal{F}(z)$ for 3\AA~$<~z~<$~5\AA.


\begin{figure*}[ht!]
\centering
\hspace{-0.3in}\includegraphics[width=7.2in]{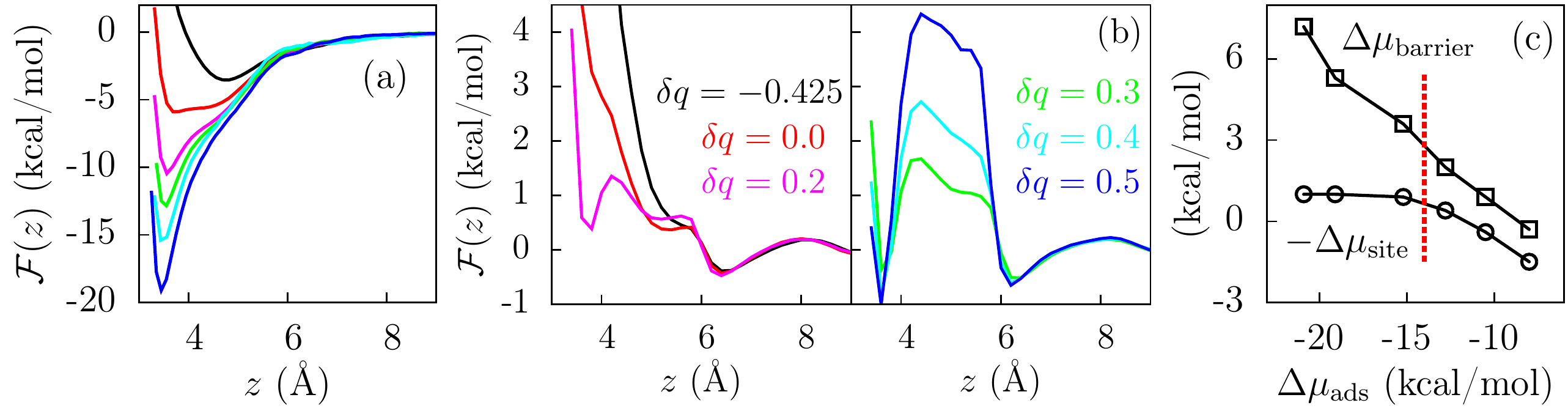}
\caption{
(a) $\mathcal{F}(z)$ for a single water on various talc surfaces 
modified to span a range of $\Delta\mu_{\rm ads}$-values. 
(b) The corresponding $\mathcal{F}(z)$ curves at saturation.
(c) The relative stability of water in the binding site compared to that in bulk, $-\Delta\mu_{\rm site}$, and the barrier to escape the binding site,
$\Delta\mu_{\rm barrier}$, as a function of the binding strength, $\Delta\mu_{\rm ads}$.
The dashed vertical line corresponds to $\mu_{\rm sat}$, the chemical potential at saturation.
} 
\label{fig:strength}
\end{figure*}


\subsection{Modified Talc Surfaces}
\label{sec:Strength}

While the H-bonding between binding sites on the talc surface and water 
leads to an interesting transition from hydrophilic at low RH to hydrophobic at 
high RH, the binding sites interact weakly with water in fluorotalc and 
pyrophyllite, which display hydrophobic behavior for all RH. To investigate the 
effect of the binding strength on the hydrophobicity of the surface, following Giovambattista 
{\it et al}.~\cite{Giovambattista_JPhysChemB_2007}, we 
construct a series of modified talc surfaces. The only force field parameters
that are changed are the charges on the oxygen (from $q_{\rm O}=-0.95$ to $q_{\rm
O}-\delta q$) and the hydrogen (from $q_{\rm H}=0.425$ to $q_{\rm H}+\delta q$) of
the hydroxyl group. We study modified talc surfaces for $\delta q$ ranging from
-0.425 which corresponds to a non-polar binding site similar to that in fluorotalc, 
to +0.6 which corresponds to an ion-pair. $\delta q=0$ is the talc surface, by definition. 

In~\ref{fig:strength}(a), we show $\mathcal{F}(z)$ for an 
isolated water molecule on the modified talc surfaces. 
As the polarity of the -OH bond is increased, the magnitude of
$\Delta\mu_{\rm ads}$ also increases, providing us with surfaces that 
display a wide range of binding strengths. 
$\mathcal{F}(z)$ at saturation, shown in~\ref{fig:strength}(b) 
for these surfaces is particularly interesting. 
For weakly adhesive surfaces ($-0.425\le\delta q<0.1$), there is only one stable 
basin at $z\approx6.5$~\AA, corresponding to molecules outside the
binding site cavity. For stronger adhesion (larger $\delta q$), a second basin 
develops at $z\approx3.5$~\AA\ and is separated from the first basin by a 
barrier. 

\ref{fig:strength}(c) shows the depth of this minimum relative to bulk,
$\Delta\mu_{\rm site}$, as a function of $\Delta\mu_{\rm ads}$. 
As the surface becomes more adhesive, more waters occupy the binding site and the depth of this minimum increases. When adhesive interactions are large enough to overcome cohesive interactions, {\it i.e.}, when $-\Delta\mu_{\rm ads}$ becomes larger than the chemical potential at saturation, $-\mu_{\rm sat}$ (for $\delta q\approx 0.4$), every binding site is occupied by a water molecule, resulting in a plateau in $\Delta\mu_{\rm site}$.

However, the height $\Delta\mu_{\rm barrier}$ of the barrier to escape 
the cavity, also shown in \ref{fig:strength}(c), continues to increase 
approximately linearly with the binding strength.
Thus, for surfaces with strong binding, $\Delta\mu_{\rm barrier}$ is large, and the exchange of molecules between the cavities and the liquid is expected to be very slow, with possible consequences on the extent of stick/slip at such surfaces in the presence of a hydrodynamic flow.

\begin{figure}[t]
\centering
\hspace{-0.15in}\includegraphics[width=3.45in]{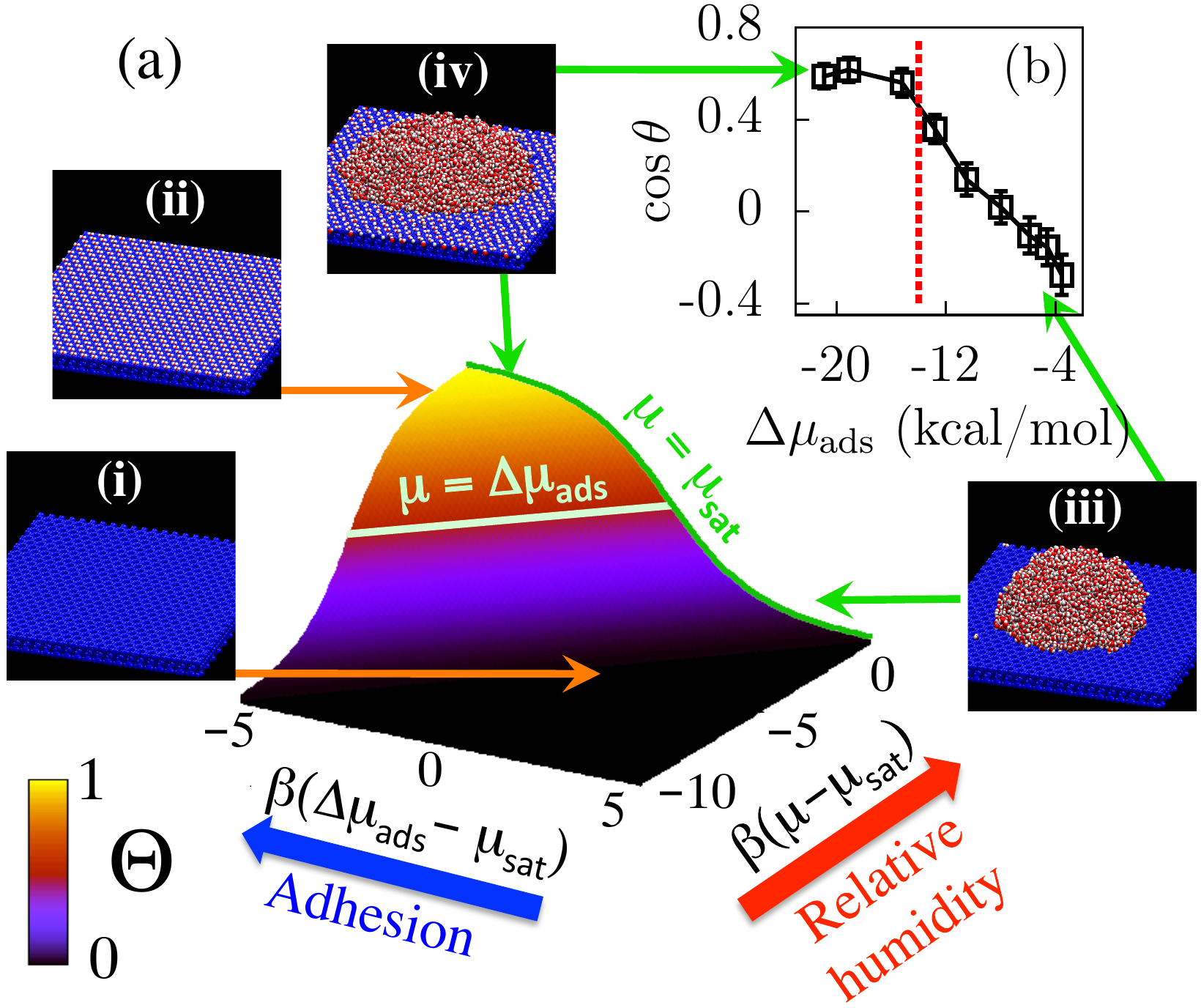}
\caption{
(a) Schematic showing the surface coverage, $\Theta$, over a wide range 
of relative humidities (${\rm RH}\equiv P/P_{\rm sat}\sim\exp[\beta(\mu-\mu_{\rm sat})]$) 
and adhesive interaction strengths ($\Delta\mu_{\rm ads}$). 
 (b) Effect of $\Delta\mu_{\rm ads}$ on surface hydrophobicity quantified by $\cos\theta$. The dashed vertical line corresponds to $\mu_{\rm sat}$.
Snapshots indicating typical configurations of water molecules (red and white) 
on modified talc surfaces (blue) are also shown. 
As the adhesive interactions ($\Delta\mu_{\rm ads}$) overcome the cohesive interactions ($\mu$), there is a transition from a dry surface [snapshots (i) and (iii)] to one covered with a monolayer of water [snapshots (ii) and (iv)].
}
\label{fig:bigpicture}
\end{figure}

\subsection{Tuning cohesion/adhesion via RH/$\Delta\mu_{\rm ads}$}
\label{sec:Conclusion}

Collectively our results paint a comprehensive picture of how the 
experimentally measurable quantities, the surface coverage $\Theta$, and the contact angle $\theta$, 
respond to changes in relative humidity  (or water chemical potential), and on the  strength of the adhesive surface-water interactions.  
The surface coverage $\Theta$, is defined as the fraction of binding sites occupied by water molecules, and its dependence on RH and $\Delta\mu_{\rm ads}$ is shown schematically in \ref{fig:bigpicture}(a).

At low RH ($\equiv P/P_{\rm sat}$), the competition between the adhesive interactions and the entropy of being in the vapor determines the surface coverage, $\Theta$.
At very low RH , there are no interactions between adsorbed waters and $\Theta$ can be approximated as~:
\begin{equation}
\Theta\approx P/P^{*}=0.1(P/P_{\rm sat})e^{-\beta\Delta\mu_{\rm ads}-8.3},
\end{equation}
where the second part of the equation is obtained by substituting for $P^{*}$ 
using Equation (1), and using appropriate values of the constants that depend on the surface geometry, $\sigma_{\rm max}$ and $\delta$, and those that depend on thermodynamic conditions, $T$~and~$P_{\rm sat}$.

For surfaces with small adhesive interactions, {\it i.e.}, $-\Delta\mu_{\rm ads}<5~$kcal/mol 
(or $-\beta\Delta\mu_{\rm ads}<8.3$), the coverage remains small ($\Theta<0.1$) 
even at saturation [snapshot (i) in~\ref{fig:bigpicture}]. Thus, no appreciable interactions between waters 
are expected over the entire range of RH-values. Both pyrophyllite 
and fluorotalc fall in this regime.

Since $\Theta$ increases exponentially with $\beta\Delta\mu_{\rm ads}$, for 
values of $-\Delta\mu_{\rm ads}>5$~kcal/mol, there can be substantial coverage 
even at modest RH [snapshot (ii) in~\ref{fig:bigpicture}]. Equation (2) is then valid only for small RH-values for 
which the predicted $\Theta$-values are small. Talc lies in this regime.

For larger RH values, there are appreciable interactions between the waters, and it is the competition between adhesive and cohesive interactions that determines surface properties. 
For surfaces such as talc, for which $-\Delta\mu_{\rm ads}<-\mu_{\rm sat}$, 
cohesion prevails at saturation, and the adsorbed waters bead up into a droplet, while the rest of the binding sites on the surface are devoid of waters [snapshot (iii) in~\ref{fig:bigpicture}]. 
Thus, the interesting crossover from hydrophobic to hydrophilic behavior in talc is a result of its adhesive interactions being strong enough to overcome vapor phase entropy at low RH, but not strong enough to overcome cohesive interactions at saturation.  
In this regime, with increasing polarity of the binding site, the surface gradually shifts from
hydrophobic to hydrophilic, and $\cos\theta$ increases approximately linearly 
as shown in \ref{fig:bigpicture}(b). 

Finally, for surfaces with even larger values of $-\Delta\mu_{\rm ads}$ that are greater than $-\mu_{\rm sat}$, adhesion dominates. . 
Surprisingly, water does not fully wet the surface at saturation. 
Instead, all binding sites are occupied by water molecules and only this first layer of water wets the surface. This water is strongly bound to the surface and the microstructure of the surface dictates the relative positions of the waters. In the present case,
the arrangement of waters on the surface is not commensurate with the hydrogen bonding 
network of water, so that water beads up on the monolayer [snapshot (iv) in~\ref{fig:bigpicture}]. For the modified talc surfaces with 
$-\Delta\mu_{\rm ads}>-\mu_{\rm sat}$, the surface has a strongly 
adsorbed water monolayer with a droplet on it that makes a contact angle of 
about $50^{\circ}$.

Similar behavior was reported by Ohler {\it et al.} for titanium dioxide surfaces, with droplet contact angles of $32-34^{\circ}$ on top of roughly two monolayers of water~\cite{Ohler_JPhysChemC_2009}.
However, other simulation studies investigating the effects of surface polarity on hydrophobicity~\cite{Giovambattista_PNAS_2009,Surblys:JCP:2011}, do not observe a plateau with non-zero contact angle at large polarities, seen in our results~[\ref{fig:bigpicture}(b)].
 Our modified talc surfaces are different from these previous studies in that the variation in polarity was achieved by changing the charges on atoms in recessed binding sites, while the remaining surface atoms remained the same. In contrast, in ref.~\cite{Giovambattista_PNAS_2009}, the surface was modified by changing dipoles that protrude from the surface, while leaving the remaining surface atoms unchanged; whereas in ref.~\cite{Surblys:JCP:2011}, the charges on all atoms in the top two layers of an FCC crystal (111 facet) were changed to tune the polarity. Thus, our results indicate that the microstructure of the surface is important in determining the effect of polarity on its wetting properties.

In contrast to the wetting properties of the model FCC surfaces used in ref.~\cite{Surblys:JCP:2011}, experimental measurements indicate that the FCC crystals of platinum (Pt), palladium (Pd), and gold (Au) are hydrophobic. 
Kimmel {\it et al.} observed a hydrophobic water monolayer on both Pt(111) and Pd(111) crystals~\cite{BruceKayPt,BruceKayPd}. 
Similarly, water has been shown to bead up on Au surfaces~\cite{Belfort:Langmuir:2010} with a contact angle of $100^{\circ}$ and Au surfaces have also been shown to adsorb, and facilitate the unfolding of proteins~\cite{Belfort:Langmuir:2011}; behavior that is typically associated with hydrophobic surfaces~\cite{AJP_Lscale}.
We speculate that the hydrophobicity of these metal surfaces arises from the presence of a monolayer of water, which binds strongly to the surface in a geometry that inhibits hydrogen bonding to the subsequent liquid water molecules.  

Our results also have implications on the wetting properties of nanoporous silicates such as hydrophobic zeolites~\cite{Cailliez_PhysChemChemPhys_2008,Cailliez_JPhysChemC_2008,Davis:PNAS} and metal-organic frameworks~\cite{Paranthaman_PhysChemChemPhys_2010}. These hydrophobic pores are thought to be devoid of water at ambient conditions, with water intrusion into the pores occurring only at sufficiently high water pressures. 
Our results suggest that in the presence of strong binding sites, these nanoporous materials may contain strongly adsorbed water molecules, even at lower pressures. If the resulting water-covered surface is hydrophobic, no further filling of the pores (analogous to wetting for planar surfaces) would be observed at ambient pressures, and intrusion would occur only at higher pressures.

\section*{Acknowledgements}
The authors thank Virginie Marry, Patrick Varilly, Mark Davis, Shekhar Garde and Adam Willard for helpful discussions. B.R. is grateful to the University of California, Berkeley, for its hospitality. A.J.P. was supported by NIH Grant No. R01-GM078102-04. D.C. was supported by the Director, Office of Science, Office of Basic Energy Sciences, Materials Sciences and Engineering Division and Chemical Sciences, Geosciences, and Biosciences Division of the U.S. Department of Energy under Contract No. DE- AC02-05CH11231.

\section*{Appendix A: Unit cells}

\begin{figure}[h]
\hspace{-0.1in}\includegraphics[width=2in]{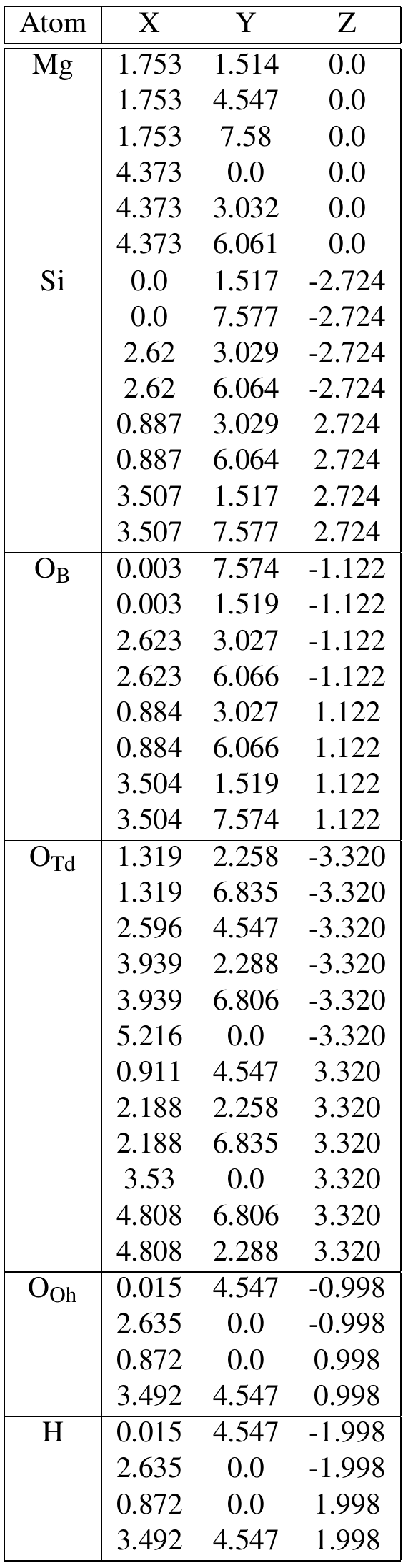}
\caption{ \label{tab:CellTalc}
Atomic coordinates in the talc unit cell.
Subscripts for oxygen differentiate tetrahedral (Td),
bridging (B) and octahedral (Oh) atoms.
}
\end{figure}

\begin{figure}[h]
\hspace{-0.1in}\includegraphics[width=2in]{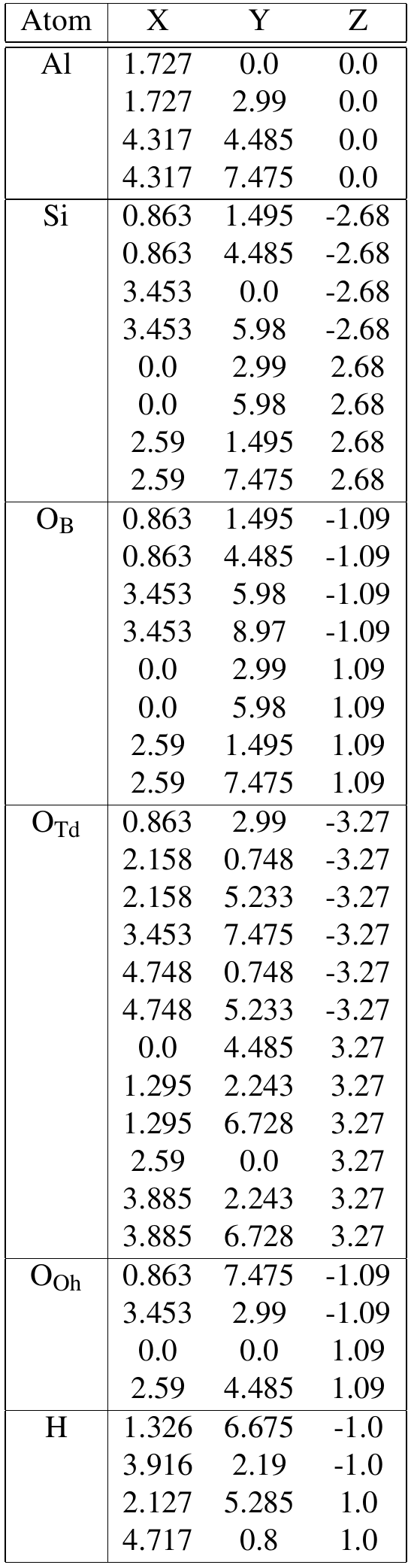}
\caption{ \label{tab:CellPyro}
Atomic coordinates in the pyrophyllite unit cell.
Subscripts for oxygen differentiate bridging (B),
tetrahedral (Td) and octahedral (Oh) atoms.
}
\end{figure}

The unit cells used for the simulation of talc and pyrophyllite
are reported in~\ref{tab:CellTalc} and~\ref{tab:CellPyro}. For fluorotalc, the oxygen of the hydroxyl group is replaced by a fluorine atom and the hydrogen is removed.
The unit cell of pyrophyllite, a dioctahedral smectite, has dimensions along
the surface of $5.18\times8.97$~\AA$^2$, as known from X-ray 
diffraction~\cite{Maegdefrau37}. The unit cell of fluorotalc is not known
exactly; we used the one determined by X-ray diffraction on synthetic 
fluorohectorite~\cite{Breu_ZanorgallgChem_2003}, which differs from fluorotalc only by substitution of some magnesium by lithium in the octahedral
layer, resulting in a permanent negative charge compensated by sodium
counterions. The unit cell has dimensions $5.24\times9.09$~\AA$^2$ along the surface. For talc we used the same structure, replacing each fluorine by a
hydroxyl group with a bond length of 1~\AA, oriented perpendicular to the surface. 

\section*{Appendix B: Comparison of force fields}

In the present work, we used the CLAYFF force field to describe the clay
surfaces and their interactions with water molecules. To justify this
choice, here we compare the predictions of another commonly used force field, 
and those of CLAYFF, with experimental results. 
This force field was originally developed by
Skipper \textit{et al.}~\cite{Skipper89} and adapted by Smith \textit{et
al.}~\cite{Smith98} for its use in conjunction with the SPC/E water model.

\begin{figure}[ht!]
\centering
\hspace{-0.1in}\includegraphics[width=3.4in]{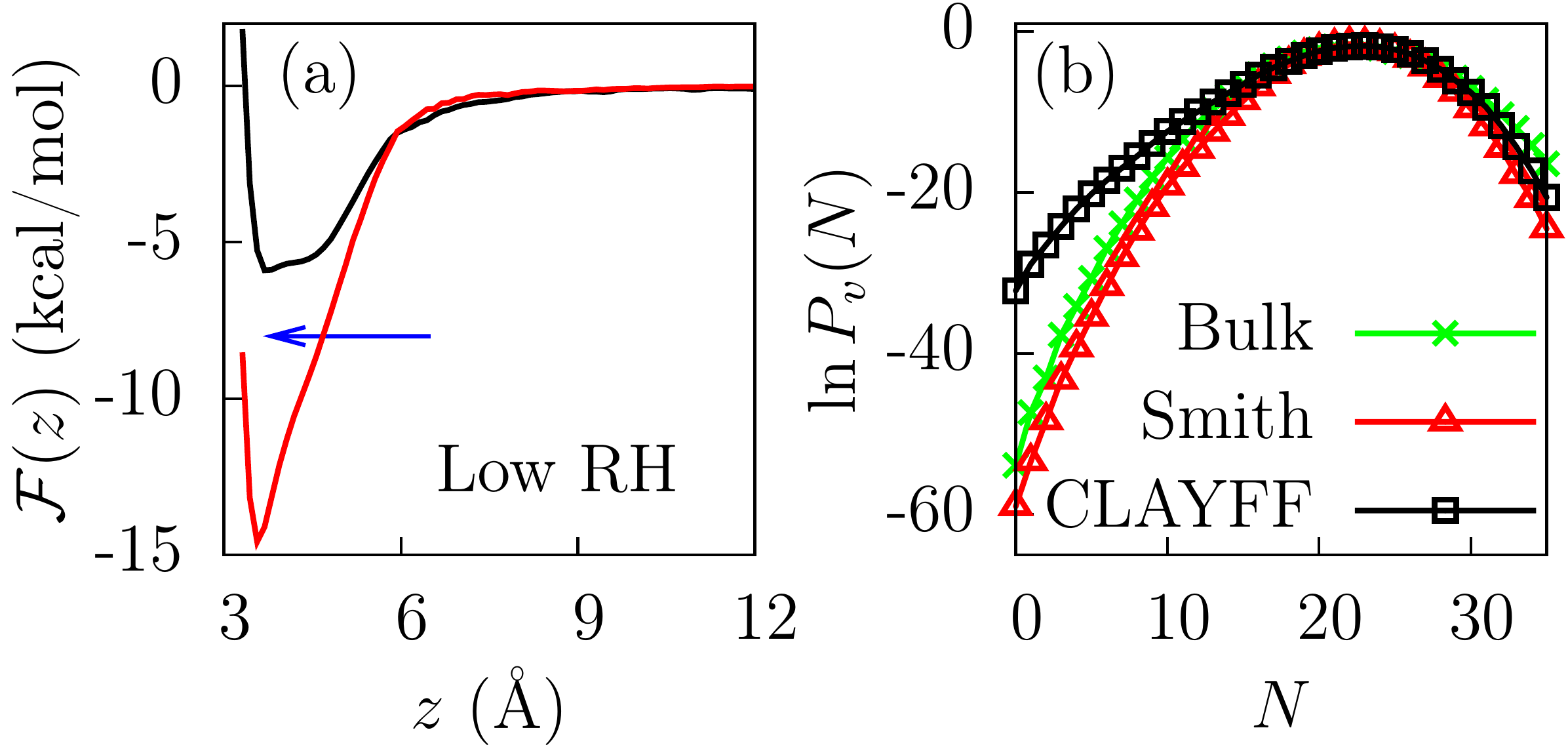}
\caption{
(a) $\mathcal{F}(z)$ for the adsorption of an isolated water molecule
on talc simulated using the CLAYFF and Skipper/Smith force fields.
The arrow indicates the experimental value of the minimum, estimated by fitting
the adsorption isotherm of Michot {\it et al.}~\cite{Michot_Langmuir_1994} 
to a Langmuir model in the very low RH regime (see text).
(b) $P_{\rm v}(N)$ for the talc surface, using the CLAYFF and Skipper/Smith force fields. 
}
\label{fig:Skipper}
\end{figure}

To investigate the talc surface at low RH, in~\ref{fig:Skipper}(a), we show the $\mathcal{F}(z)$ obtained using the Skipper/Smith force field and compare it with that obtained using the CLAYFF force field. Also shown is the experimental estimate discussed in the main text, indicating that the Skipper/Smith force field overestimates the binding or adsorption free energy. 

To investigate the hydrophobicity of talc surfaces at saturation, obtained using the two force fields, in~\ref{fig:Skipper}(b), we show the respective $P_{\rm v}(N)$ distributions. 
$P_{\rm v}(N)$ for $\mathrm{v}$ near the Skipper/Smith talc surface indicates that it is harder to empty the observation volume close to the surface than in bulk water. This is also consistent with the observed complete wetting of the talc surface by a droplet, 
indicating a contact angle of $\theta=0^\circ$. Such a complete wetting is
however in contradiction with the experimental contact angle of $80-85^\circ$.
We thus conclude that the Skipper/Smith force field significantly overestimates talc-water adhesive interactions, both at low RH and at saturation.

Another force field used to model dioctahedral clays and their interaction with organic
cations was proposed by Heinz \textit{et al.}~\cite{Heinz_ChemMater_2005}.
This model was not extended to triocahedral clays such as talc,
and the behavior of water at clay surfaces modeled with this
force field has not been reported. We nevertheless simulated water
droplets on the surface of pyrophyllite using this force field.
The resulting contact angle ($125^\circ$) was larger than that measured 
experimentally ($80-85^\circ$), suggesting that this force fields results in 
surfaces that are too hydrophobic. 

Finally, while we find that CLAYFF is the best available force field to date, to simulate water at the surface of uncharged clay minerals, the present work suggests that it is too hydrophobic. Thus we find that there is room for improvement to describe the clay-water interaction, in agreement with the findings of a recent study comparing molecular simulations with  X-ray and neutron diffraction experiments on a charged smectite~\cite{Ferrage_JPhysChemC_2011}. The insights gained during the present study of neutral clays, which are more sensitive to the clay-water interactions, could also be helpful in the design of an improved force field. Such design requires a subtle balance between different interactions which is generally not achieved by tuning only one parameter. With this caveat in mind, we note that a slightly more polar hydroxyl group might be relevant, as the modified talc surface with  $\delta q=0.1$ seems to agree quite well with experimentally measured $\Delta\mu_{\rm ads}$ and $\cos\theta$ values for talc.



\providecommand*{\mcitethebibliography}{\thebibliography}
\csname @ifundefined\endcsname{endmcitethebibliography}
{\let\endmcitethebibliography\endthebibliography}{}


\providecommand{\refin}[1]{\\ \textbf{Referenced in:} #1}


\begin{mcitethebibliography}{60}
\bibitem[edg()]{edges}
 Since the number of hydrophilic binding sites on clay edges is much smaller
  than on the talc surface, we neglect the edge sites to obtain this
  estimate.\relax
\mciteBstWouldAddEndPunctfalse
\mciteSetBstMidEndSepPunct{\mcitedefaultmidpunct}
{}{\mcitedefaultseppunct}\relax
\EndOfBibitem
\end{mcitethebibliography}
\end{document}